\documentclass[12pt,a4paper,oneside]{article}
\usepackage{graphicx}

\begin{document}

\title{Stabilizing single atom contacts by molecular bridge
formation} \maketitle

\author{Everardus H. Huisman*\dag, Marius L. Trouwborst\dag, Frank L. Bakker\dag, Bert de Boer\dag, Bart J. van Wees\dag, Sense J. van der
Molen\ddag}

\emph{Zernike Institute for Advanced Materials,University of
Groningen, Nijenborgh 4, 9747 AG Groningen, The Netherlands, and
Kamerlingh Onnes Laboratory, Leiden University, Niels Bohrweg 2,
2333 CA Leiden, The Netherlands} \footnote{\dag University of
Groningen \ddag University of Leiden and University of Groningen,
*Corresponding author email address: e.h.huisman@rug.nl}

\maketitle

\begin{abstract}
Gold-molecule-gold junctions can be formed by carefully breaking a
gold wire in a solution containing dithiolated molecules.
Surprisingly, there is little understanding on the mechanical
details of the bridge formation process and specifically on the
role that the dithiol molecules play themselves. We propose that
alkanedithiol molecules have already formed bridges between the
gold electrodes \textit{before} the atomic gold-gold junction is
broken. This leads to stabilization of the single atomic gold
junction, as observed experimentally. Our data can be understood
within a simple spring model.
\end{abstract}

A single molecule forms a potential electronic component, offering
the perspective of true bottom up engineering of nanodevices.
Functionalities such as switching \cite{KATSONIS} \cite{RIEL} and
rectifying \cite{WEBER} have been demonstrated in the past years.
Nevertheless, the field of molecular electronics has been troubled
by difficulties in making reliable and well-defined contacts to
single molecules. Fortunately, recent times have seen a significant
growth of independent techniques to contact single molecules or
small ensembles of molecules \cite{TAO3} \cite{AKKERMAN}. An
important contribution to this development was made by Xu and Tao,
who used a scanning tunneling microscope (STM) to contact
dithiolated molecules \cite{TAO}. In these experiments, a gold
STM-tip is carefully pulled out of contact with a gold substrate in
the presence of a solution. Simultaneously, the conductance $G$ of
the tip-substrate junction is measured. As the tip is pulled up, the
diameter of the gold neck connecting tip and substrate becomes
smaller and hence $G$ decreases. This process is continued down to
the limit of a single gold atom contact, in which case $G$ reaches a
value around $1 G_0 = 2e^2/h=77.5 \mu$S, the quantum of conductance.
Pulling further, one observes a sudden downward jump in the
conductance, indicative of the breaking of the gold constriction.
This 'jump out of contact' (JOC) has been thoroughly studied by
several groups \cite{AGRAIT,TROUWBORST}. Xu and Tao made the
remarkable observation that a gold-molecule-gold contact may form
after JOC, provided molecules with suitable anchor groups (such as
thiol groups) are present in the solution \cite{TAO}. The molecular
junctions thus created have a limited life time and a conductance
that varies from experiment to experiment. Therefore, it is pivotal
to carefully study the statistics of many of such 'break junction'
traces. Several groups have adopted the Tao method since, either
using scanning tunneling microscopy break junctions (STMBJ)
\cite{LINDSAY,LI,HAISS,VENKATARAMAN}, or mechanically controllable
break junctions (MCBJ) \cite{GONZALEZ}. Although the procedure is
applied to a variety of molecules \cite{ KAWAI2,VENKATARAMAN2,TAO4},
many studies have focused on simple alkanedithiol molecules in order
to create a well-defined reference point for molecular electronics
\cite{LINDSAY,LI,HAISS, VENKATARAMAN,GONZALEZ,FUJIHARA,MARTIN,
AGRAIT2}. Despite these efforts, remarkably little is known about
the mechanism of molecular bridge formation. Here, we will address
this important, but relatively untouched issue.

\begin{center}
\begin{figure}[p]
  \includegraphics[width=13.5cm]{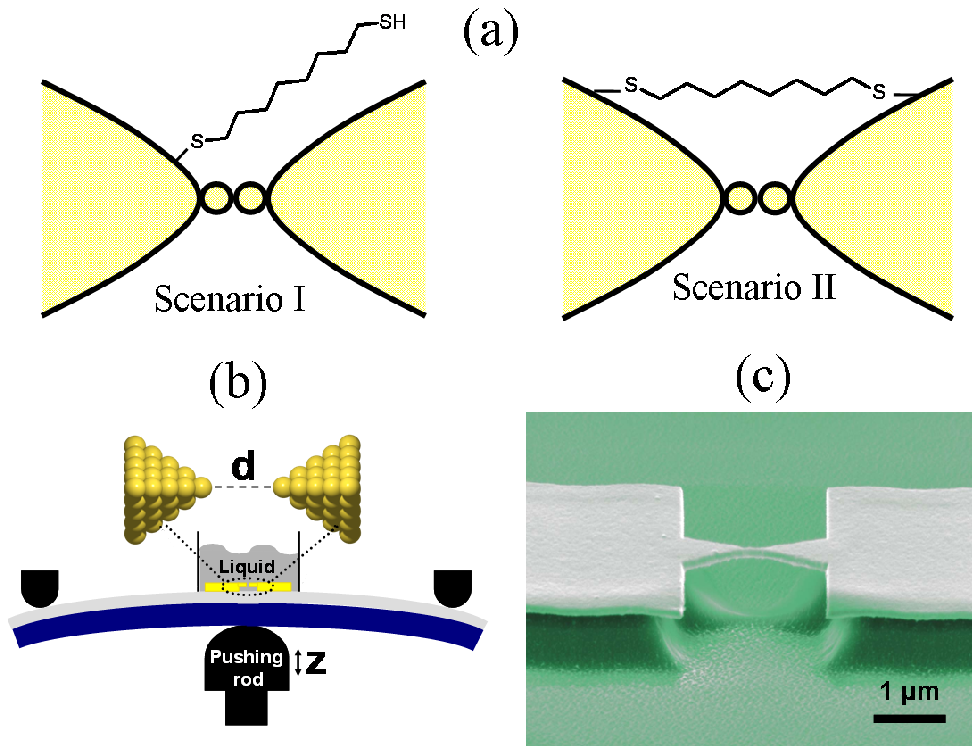}\\
\caption{\textbf{(a)} Two possibilities for the position of a
dithiolated molecule just before breaking of the gold wire. In I,
the molecule is attached to one side of the electrode only.  In II,
the molecule is attached to both electrodes. \textbf{(b)} Schematic
drawing of the MCBJ technique showing the liquid cell on top of the
microfabricated gold leads on the flexible substrate clamped in a
three-point bending configuration. \textbf{(c)} Scanning electron
micrograph showing the suspended gold bridge on top of the polyimide
layer.} \label{Figure1}
\end{figure}
\end{center}

On our quest, we focus on the moment right before the jump out of
contact. At that instance, the dithiol molecule(s) that will later
form the bridge are either connected to one (scenario I in Figure
\ref{Figure1}(a)) or to both electrodes (scenario II). A situation
in which no dithiol molecules are connected to the gold contacts
prior to breaking is highly unlikely due to the strong tendency of
Au-S bond formation. Interestingly, both scenario I and II can
evolve into a metal-molecule-metal bridge. In I, the
metal-molecule-metal bridge is formed after JOC, when the loose end
of the molecule binds to the other electrode. In II, the molecule
already bridges both sides of the electrodes before JOC. The
presence of such molecular bridges would not be obvious from the
conductance value itself, due to the much higher parallel current
flowing through the gold neck. However, the atomic junction would be
reinforced by parallel molecular bridges, leading to a higher
mechanical stability of the gold constriction. After the atomic
junction breaks down, the conductance of a molecular bridge can
finally be determined. In this Letter, we provide evidence for
scenario II. Our experimental results are discussed in the light of
a simple spring model.

For our experiments, we use lithographically defined MCBJs,
submerged in a toluene solution \cite{GONZALEZ, GRUTER}. A schematic
representation of the set-up is given in Figure \ref{Figure1} (b).
In short, a MCBJ consists of a gold wire with a constriction in its
center, attached to a flexible substrate (see Figure \ref{Figure1}
(b,c)) \cite{AGRAIT}. The substrate is bent in a 3-point bending
mechanism, by moving a pushing rod upwards with a motor. As a
result, the central constriction is gently elongated, until it
finally breaks. After this, the distance between the two freshly
created electrodes, $d$, is related to pushing rod position, $Z$, by
the attenuation factor, $r=\triangle d / \triangle
z$=$\zeta6ut/L^2$. Here $u$ is the length of the bridge, $L$ is the
distance between the counter supports and t is the thickness of the
substrate. For lithographic MCBJs, $r$ should incorporate a factor
$\zeta$ to correct for the presence of a soft (polyimide) layer
\cite{VROUWE}. To fabricate such MCBJs, a gold (99.99\%, Umicore)
wire of 100 nm wide and 120 nm thick (using a 1 nm Cr adhesion
layer) is thermally evaporated on top of a pyrralin polyimide coated
phosphor bronze substrate (22.5 $\times$ 10.5 $\times$ 0.5 mm) with
standard lift-off based e-beam lithography. The central constriction
is made free hanging by reactive ion etching with an O$_2$/CF$_4$
plasma (see Figure \ref{Figure1} (c)). Calibrating the junctions in
argon we find, $\zeta$= 4.5 and r $\approx 7.7 \cdot10^{-5}$ for our
specific geometry ($u=2.7 \mu$m, $L=20$ mm, $t=0.42$ mm). Hence, d
can be controlled with sub-\AA$ $ resolution, with a drift of less
than 1 pm/min at room temperature. The conductance of the junction
is measured by applying a 100 mV bias, while sampling the current at
5 kHz with a 16-bit National Instruments data acquisition board via
a home-built trans impedance amplifier (1$\mu$A/V) \cite{GONZALEZ,
GRUTER}. We add a series resistance to limit the total current
(101.3k$\Omega$) at low junction resistances (thereby decreasing the
effective bias felt by the junction). As a solvent, we choose
nitrogen-saturated toluene (see Supporting Information) due to its
good solubility for organic molecules, low conductivity and low
hygroscopy. Alkanedithiol molecules, i.e., 1,4-butanedithiol (BDT)
and 1,8-octanedithiol (ODT), were obtained from Sigma-Aldrich.

After mounting the sample and the liquid cell, we flush with 40 ml
of toluene. Subsequently, we introduce 20 ml of the solution of
interest, which is either pure toluene or toluene with a
concentration of 10 mM alkanedithiols. Next, we start breaking and
rejoining the gold junction by moving the pushing rod up and down.
Traces of conductance versus pushing rod position, $G(Z)$, are
recorded with a pushing rod speed of +15 $\mu$m/s, corresponding to
a local elongation of 1.1 nm/s. After having reached the lowest
measurable current (just above 10$^{-11}$ A), we push another 30
$\mu$m in Z in order to allow diffusion of molecules in between the
electrodes. Then, the junction is closed again to a conductance
value of $\sim$ 10 $G_{0}$ in order to randomize the atoms and
molecules involved.

\begin{center}
\begin{figure}[p]
  \includegraphics[width=15cm]{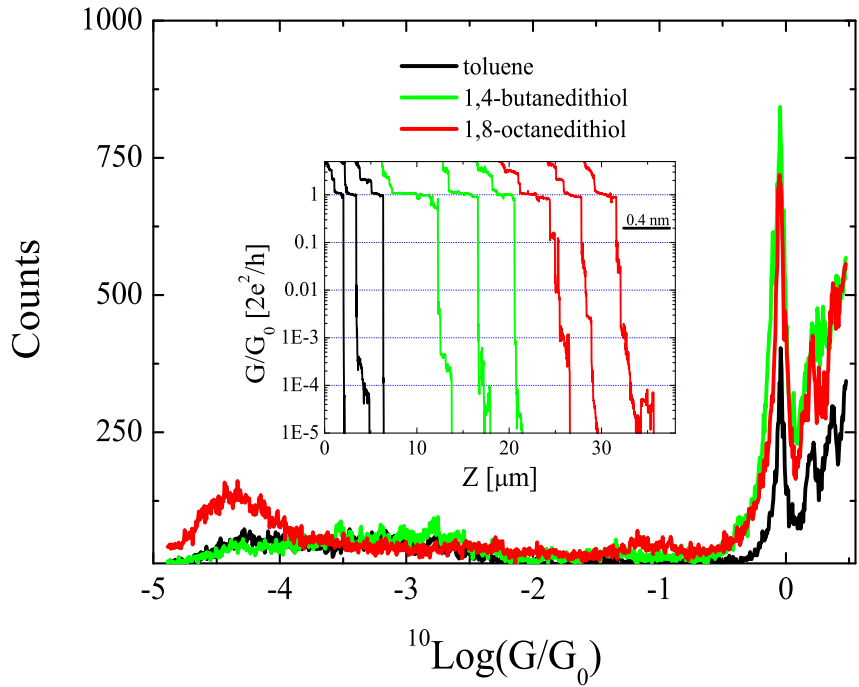}\\
\caption{Logarithmic conductance histogram of 250 opening traces in
toluene of sample TOL1 (black), in 10 mM BDT of sample BDT1 (green)
and in 10 mM ODT of sample ODT1 (red). The conductances values were
collected in bins of 0.0054 $^{10}Log(G/G_0)$. The inset displays 3
sample traces for each solution. The scale bar shows the actual
electrode displacement $d$. The data has been corrected for an
effective series resistance of 490 $\Omega$ \cite{COSTA}. }
\label{Figure2}
\end{figure}
\end{center}

The inset of Figure \ref{Figure2} displays 3 typical opening traces
on a semilog plot in pure toluene, i.e., without dithiols (black).
While breaking gold nanowires, we distinguish 2 different regimes:
the contact regime and the out-of-contact regime. In the contact
regime the conductance is given by the Landauer formula: $G =
G_0\sum{T_n}$, where $T_n$ is the transmission probability of the
n$^{th}$ channel. A single gold atom acts as a waveguide for
electrons and forms a single channel \cite{AGRAIT,BEENAKKER, KAWAI}.
In our experiments this is especially visible just before breaking,
when only one atom bridges the electrodes. Due to the stability of
this conformation, plateaus at a constant conductance value around 1
$G_0$ appear. This plateau abruptly ends by a JOC to lower
conductances (the out-of-contact regime) within 1 ms. When the
junction breaks for the first time, the conductance after JOC drops
to values below 10$^{-5}$ $G_0$ (the first black trace in the inset
of Figure \ref{Figure2}). Usually, within tens of traces the
conductance just after JOC increases to values around $1
\cdot$10$^{-3}$ $G_0$. At low temperatures, we have previously
showed that the size of JOC can be controlled by 'training' the
contact \cite{TROUWBORST}. This procedure reduces the number of
atoms involved in the breaking, eventually till the ultimate limit
of 2 atoms. We have explored the possibility of 'training'
electrodes at room temperature, thereby reducing JOC. We found that
the high mobility of gold atoms at elevated temperatures make this
procedure cumbersome. After JOC, $G$ decays roughly exponentially
with $d$ as is expected for tunneling. Using the fact that the
tunneling decay constant in toluene is roughly the same as in vacuum
(1 dec/\AA) \cite{GRUTER}, we can relate the actual jump in
conductance to a electrode distance of about 3 \AA.

In the main panel of Figure \ref{Figure2}, a histogram for pure
toluene (black, solid line) incorporating all 250 opening traces in
1000 logarithmic bins is shown.  Such a representation, allows for a
broad overview of the entire data set, while correcting for
background tunneling in a natural way \cite{GONZALEZ}. We emphasize
that we did not select traces. In the contact regime, the histogram
shows peaks at (integer values of) 1 $G_0$, as will be discussed in
more detail below. In the out-of-contact regime, the histogram is
relatively featureless, except for the lack of points in the regime
where the jump out of contact takes place ($1G_0>G>3 \cdot 10^{-3}
G_0$). Most importantly, we note that for $G<3\cdot 10 ^{-3} G_0$,
the number of points per bin is roughly constant. This is a result
of the exponential decay in $G$ due to tunnelling in pure toluene.

Next, we repeat the experiment in the presence of ODT molecules (10
mM in toluene). In that case, additional plateaus of constant
conductance appear in the out-of-contact regime in about 20 \% of
the opening traces (red traces in the inset of Figure
\ref{Figure2}). These plateaus are usually shorter than the ones at
1 $G_0$ and show fluctuations in the conductance. They are
interpreted as the signature of gold-molecule-gold bridges. To
obtain a statistically sound value for the conductance of an ODT
molecular bridge, we collect all 250 traces in a logarithmic
histogram (red line in the main panel of Figure \ref{Figure2}). The
conductance plateaus discussed above, result in a peak around
$G=4\cdot$ 10$^{-5}$ $G_0$, related to molecular bridge formation.
The average conductance we find for ODT bridges is in good agreement
with the work by Gonz\`alez \emph{et al.}, which was done under
similar circumstances \cite{GONZALEZ}. Furthermore, it is in
correspondence with the so-called 'low peaks' in the work of the Tao
group \cite{LINDSAY} and the 'medium peaks' in the work of the
Wandlowski group \cite{LI}. When adding a shorter molecule, i.e.,
BDT, an interesting contrast is found with the case of ODT. For BDT,
no additional peaks appear in the out-of-contact regime (green
traces and green histogram in Figure \ref{Figure2}).

\begin{center}
\begin{figure}[p]
  \includegraphics[width=15cm]{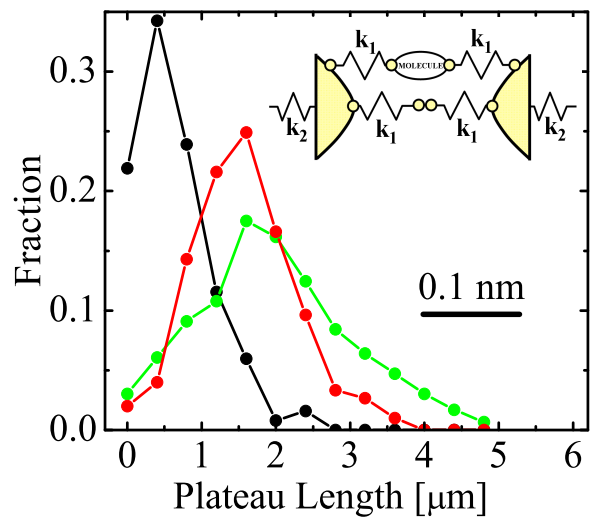}\\
\caption{Plateau length histogram of 250 opening traces in toluene
of sample TOL1 (black), 300 opening traces in 10 mM BDT of sample
BDT1 (green) and 300 opening traces in 10 mM ODT of sample ODT1
(red). The bins are linear in units of $Z$. The plateau length is
defined as the length (in units of $Z$) of the plateau between 0.5
and 1.5 $G_0$. The inset shows a simple spring model to explain the
increase in $G_0$ plateau length. The model incorporates alkane
dithiols as parallel bridges in accordance with scenario II in
Figure \ref{Figure1}(a).} \label{Figure3}
\end{figure}
\end{center}

Having discussed the out-of-contact regime, where the conductance
values of molecular bridges can be directly distinguished, we focus
on another remarkable difference between toluene and dithiol
experiments. For this, we concentrate on the contact regime, i.e.,
before the atomically thin gold neck is broken. The formation of
single or few atomic gold junctions during breaking gives rise to
plateaus around (multiples of) $G_0$ (inset Figure \ref{Figure2}).
In histograms, this translates to peaks (main panel of Figure
\ref{Figure2}). Interestingly, these peaks are significantly larger
in the presence of BDT and ODT than for pure toluene. This effect
(both in peak height and area) is especially clear around 1 $G_0$. A
first, rather trivial possibility would be that the addition of
dithiols to toluene may lead to a decrease in the attenuation factor
$r$. A smaller $r$ would give rise to a lower local velocity $\Delta
d/ \Delta t$ and therefore to more points per bin. This effect
should be especially clear in the tunnel regime, which is very
sensitive to variations in distance. In a logarithmic
representation, a lower $r$ should yield a higher constant
background in the out-of-contact regime \cite{GONZALEZ}. However,
besides the ODT peak around 4 $\cdot$ 10$^{-5}$ $G_0$, no apparent
increase in counts is observed in the out-of-contact regime for both
ODT and BDT (say for $10^{-4}G_0<G<3 \cdot 10^{-3} G_0$). This rules
out a variation of $r$, indicating that the presence of dithiols
truly stabilizes atomic contacts. To substantiate this, histograms
of the length of the $G_0$ plateaus were constructed for all
experiments (Figure \ref{Figure3}). The plateau length is defined as
the length (in units of $Z$) of the plateau between 0.5 and 1.5
$G_0$. We summarized the average plateau length of 7 different data
sets (3 times pure toluene, 2 times 10 mM BDT and 2 times 10 mM ODT)
in Table \ref{TABLE1}. Clearly, the effect reproduces using
different samples and the 1 $G_0$ plateau length shifts towards
higher values in going from toluene (average value of 0.7 $\mu$m) to
ODT and BDT (1.6 $\mu$m and 2.0 $\mu$m, respectively). Note that
these plateau length values correspond to a displacement close to
the diameter of a gold atom ($\frac{0.25 nm}{r} = 3.3 \mu$m),
indicating that no chains of atoms are formed during the breaking
procedure \cite{ALEX}. We conclude that, in the presence of
dithiols, the electrodes have to be displaced over a larger distance
to break the gold-gold junction. In other words, alkanedithiols
reinforce the atomic gold junction. This is fully consistent with
scenario II (Figure \ref{Figure1}(a)), in which molecular bridges
have already formed before the gold neck breaks.

To describe how the presence of molecular bridges leads to longer
$G_0$ plateaus in $G(Z)$-curves, we present a tentative model, which
is depicted in the inset of Figure \ref{Figure3}. Here, we have
translated scenario II into a simple spring model
\cite{TROUWBORST,TORRES,OLESEN}. The atomic contact itself will
generally be a gold dimer \cite{TROUWBORST,UNTIEDT}. The role of the
$n$ molecular bridges is to form $n$ parallel springs, strengthening
the junction. The attachment of the dimer to the first atomic layers
of each of the electrodes can be described by a spring $k_1$
\cite{TROUWBORST}. The elastic properties of the remainder of the
electrode (the bulk of the electrode) can also be described by a
spring $k_2$. This description is similar to that of Torres \emph{et
al.}, where the contact is modeled as a series of $N$ slices with a
spring, $k_n$ \cite{TORRES}. As for the molecular bridges, we assume
that they are rigid, i.e. all displacements take place in between
gold atoms \cite{TAO2}. We note that the Au-S bond is stronger than
the Au-Au bond itself \cite{TAO5}. Hence, the weakest link of the
molecular bridge is formed by the very gold atom that binds to the
molecular S atom. This gold atom is itself attached to the first
atomic layers of the electrode by a spring, with spring constant $k
\approx k_1$, which is again attached to the remainder of the
electrode. In total, we have $n+1$ identical springs (1 due to the
dimer, $n$ due to the dithiols), which are in turn attached to the
bulk electrode spring $k_2$. The total spring constant equals
$k_{tot}=\frac{a(n+1)}{a(n+1)+1}k_2$, where $a=k_1/k_2$. To break
the dimer, the two gold atoms should be pulled apart by a force $F_0
\approx 1.5$nN \cite{RUBIO}. In the absence of dithiol molecules,
this happens after the pushing rod has traveled a critical distance
$Z_0$. However, in the presence of $n$ parallel springs, a greater
total force $F_n$ is to be applied over the junction, i.e.,
$F_n=F_0(n+1)$. This force is also felt by springs $k_2$, which are
consequently elongated extra. Hence, the pushing rod has to be
pushed further, over a distance $Z_n$ to finally break the dimer.
Our simple model yields $\frac{Z_n}{Z_0}=\frac{a(n+1)+1}{a+1}$.

From the experiment (see Table \ref{TABLE1}), we find that the
increase in plateau length with respect to pure toluene is
consistently more pronounced for BDT and ODT data sets with
$\frac{Z_n}{Z_0}\approx$ 2-3. Olesen \emph{et al.} have shown that
when making contact with a metal STM tip to a metal surface,
approximately 1/4 of the initial displacement takes place between
the last atom of the tip and the first layer of the surface metal
atoms \cite{OLESEN}. The other 3/4 takes place in the neighboring
metal layers. Applying these numbers, such that $a\approx 3$, we get
a consistent picture in which a few (typically 1 to 3) molecules
bridge the atomic junction before it breaks ($\frac{Z_1}{Z_0}=7/4$,
$\frac{Z_2}{Z_0}=10/4$ and $\frac{Z_3}{Z_0}=13/4$).

After the atomic junction breaks, the metal-molecule-metal bridges
become observable in the conductance, as seen for ODT. Remarkably,
the stabilization effect in the contact regime is also observed for
BDT (Figure \ref{Figure3}), while no peak in the out-of-contact
regime appears (Figure \ref{Figure2}). This indicates that BDT
bridges break during JOC. Most likely, this is due to an 'avalanche'
effect: a sudden strain relief at JOC disrupts all junctions
present. Short molecular bridges, such as gold-BDT-gold, are likely
to bridge both electrodes in a stretched conformation, without
so-called gauche defects. Such a conformation is rigid and is
unlikely to accommodate a sudden distance jump of a few \AA. Long
molecular bridges, such as gold-ODT-gold, are less rigid, especially
when they are in a bent conformation due to gauche defects
\cite{AKKERMAN,LI}. Therefore, they are less likely to be disrupted
by JOC and explain why ODT does show plateaus in the out-of-contact
regime.

We are not aware of any conductance measurements using forced
gold-gold contact which show formation of metal-molecule-metal
bridges of alkanedithiols smaller than 1,6-hexanedithiol
\cite{LINDSAY}. However, a slightly alternative approach was
reported by Li \emph{et al.} and Haiss \emph{et al.}
\cite{LI,HAISS,GABOR,HAISS2}. Here, gold-gold contact was avoided
and electrodes with relatively low alkanedithiol coverage were used.
In this way, alkanedithiol junctions as small as 1,5-pentanedithiol
were measured and a strong preference for single molecule junctions
was observed. Li \emph{et al.} could even distinguish different
conformations and couplings of a \emph{single} molecular bridge. Our
simple spring model helps to qualitatively understand the
differences in data quality found in literature. In the experiments
of Li \emph{et al.} and Haiss \emph{et al.} no gold bridge is
present. Therefore, molecular junctions are not disrupted by the
strain release during JOC, such that clear signals of (short)
alkanedithiol bridges are observed.

In summary, we discuss new aspects of molecular bridge formation in
break junction experiments in solutions of dithiolated molecules. We
find that, in order to break a single atom contact, the electrodes
have to be displaced 2-3 times longer in the presence of
alkanedithiols as compared to the displacement when only solvent is
present. This observation provides evidence for a scenario in which
a few molecules already span the atomically thin gold neck before
breaking, thereby reinforcing the atomic contact. Although present
before JOC, metal-molecule-metal bridges only become 'observable' in
the conductance when the metal bridge breaks. Our data are supported
by a simple spring model. We put forward an important notion: A
molecular bridge should be able to accommodate the strain release
upon JOC in order to form a stable metal-molecule-metal bridge. For
alkanedithiols at room temperature we find that the cross-over is
between butanedithiol and octanedithiol.

\paragraph{Acknowledgements}
The authors acknowledge Bernard Wolfs and Siemon Bakker for
technical support, Maarten Smid for his assistance in calibrating
the setup and Oetze Staal for supplying purified toluene. This work
was financed by the Netherlands Organisation for Scientific
Research, NWO, and by the Zernike Institute for Advanced Materials.

\newpage

\begin{table}[h]
\small
\begin{center}
\begin{tabular}{lccccccccc}
    \hline
    Sample&TOL1&TOL2&TOL3&BDT1&BDT2&ODT1&ODT2\\
    \hline
    Number of traces & 250&250&150&300&50&300&150\\
    Mean plateau length ($\mu$m)  &0.62&0.74&0.65&2.02&2.01&1.55&1.58 \\
    \hline
  \end{tabular}
\caption{Table summarizing the number of traces and the mean plateau
length of 7 different samples. The plateau length for each trace is
defined as the length (in units of $Z$) of the plateau between 0.5
and 1.5 $G_0$. The mean plateau length was determined by dividing
the sum of all plateau lengths by the number of traces.}
\label{TABLE1}\end{center}

\end{table}


\begin{thebibliography}{99}

\bibitem{KATSONIS}
Katsonis, N.; Kudernac, T.; Walko, M.; van der Molen, S.J.; van
Wees, B.J.; Feringa, B.L. \emph{Adv. Mat.} \textbf{2006}, 18,
1397-1400.

\bibitem{RIEL}
L\"ortscher, E.; Ciszek, J.W.; Tour, J.; Riel, H. \emph{Small}
\textbf{2006}, 2, 973-977.

\bibitem{WEBER}
Elbing, M.; Ochs, R.; Koentopp, M.; Fischer, M.; Von H\"anisch, K.;
Weigend, F.; Evers, F.; Weber, H.B.; Mayor, M. \emph{Proc. Natl.
Acad. Sci.} \textbf{2005}, 102, 8815-8820.

\bibitem{TAO3}
Chen, F.; Hihath, J.; Huang, Z.; Li, X.; Tao,
N.J. \emph{Annu. Rev. Phys. Chem.} \textbf{2007}, 58, 535-564.

\bibitem{AKKERMAN}
Akkerman, H.B.; de Boer, B. \emph{J. Phys.: Condens. Matter}
\textbf{2008}, 20, 013001.

\bibitem{TAO}
Xu, B.; Tao, N.  \emph{Science} \textbf{2003}, 301,
1221-1223.

\bibitem{AGRAIT}
See for a review: Agrait, N.; Yeyati, A.L.; van Ruitenbeek, J.M.
\emph{Physics Reports} \textbf{2003}, 377, 81-279 and references
therein.

\bibitem{TROUWBORST}
Trouwborst, M.L.; Huisman, E.H.; Bakker, F.L.; van der Molen, S.J.;
van Wees, B.J. \emph{Phys. Rev. Lett.} \textbf{2008}, 100, 175502.

\bibitem{LINDSAY}
Li, X.; He, J.; Hihath, J.; Xu, B.; Lindsay, S.M.; Tao, N.J.
\emph{J. Am., Chem., Soc.} \textbf{2006}, 128, 2135-2141.

\bibitem{LI}
Li, C.; Pobelov, I.; Wandlowski, Th.; Bagrets, A.; Arnold, A.;
Evers, F. \emph{J. Am. Chem. Soc.} \textbf{2008}, 130, 318-326.

\bibitem{HAISS}
Haiss, W.; Nichols, R.J.; Van Zalinge, H.; Higgins, S. J.; Bethell,
D.; Schriffin, D.J. \emph{Phys. Chem. Chem. Phys} \textbf{2004}, 6,
4330-4337.

\bibitem{VENKATARAMAN}
Venkataraman, L.; Klare, J.E.; Tam, I.W.; Nuckolls, C.; Hybertsen,
M.S.; Steigerwald, M.L. \emph{Nano Lett.}, \textbf{2006}, 6,
458-462.

\bibitem{GONZALEZ}
Gonz\`alez, M.T.; Wu, S.; Huber, R.; van der Molen, S.J.;
Sch\"onenberger, C.; Calame, M. \emph{Nano Lett.}, \textbf{2006}, 6,
2238-2242.

\bibitem{KAWAI2}
Tsutsui, M.; Shoji, K.; Morimoto, K.; Taniguchi, M.; Kawai, T.
\emph{Appl. Phys. Lett}, \textbf{2008}, 92, 223110 (3 pp).

\bibitem{VENKATARAMAN2}
Venkataraman, L.; Klare, J.E.; Nuckolls, C.; Hybertsen, M.S.;
Steigerwald, M.L. \emph{Nature}, \textbf{2006}, 442, 904-907.

\bibitem{TAO4}
Xu, B.; Zhang, P.; Li, X.; Tao, N. \emph{Nano Lett.}, \textbf{2004},
6, 1105-1108.

\bibitem{FUJIHARA}
Suzuki, M., Fuji, S.; Fujihara, M. \emph{Jap. J. Appl. Phys.}
\textbf{2006}, 45, 2041-2044.

\bibitem{MARTIN}
Martin, C.A.; Ding, D.; Van der Zant, H.S.J.; van Ruitenbeek, J.M.
\emph{New J. Phys} \textbf{2008}, 10, 065008.

\bibitem{AGRAIT2}
Hihath, J.; Arroyo, C. R.; Rubio-Bollinger, G.; Tao, N.J. Agra\"it,
N. \emph{Nano Lett.} \textbf{2008}, 8 , 1673-1678.

\bibitem{GRUTER}
For a detailed description of the setup used here, see supplementary
information. Also see: Gr\"{u}ter, L.; Gonz\`alez, M.T.; Huber, R.;
Calame, M.; Sch\"onenberger, C. \emph{Small} \textbf{2005}, 1,
1067-1070.

\bibitem{VROUWE} Vrouwe, S.A.G.; van der Giessen, E.;
van der Molen, S.J.; Dulic, D.; Trouwborst, M. L.; van Wees, B.J.
\emph{Phys. Rev. B. }\textbf{2005}, 71, 35313-35319.

\bibitem{BEENAKKER}
van Houten, H.; Beenakker, C. \emph{Physics Today} \textbf{1996},
July, 22-29.

\bibitem{KAWAI}
Tsutsui, M.; Shoji, K.; Taniguchi, M.; Kawai, T.  \emph{Nano Lett.},
\textbf{2008}, 8, 345-349.

\bibitem{ALEX}
Yanson, A.I.; Rubio Bollinger, G.; van den Brom, H.E.; Agra\"it, N.;
van Ruitenbeek, J.M. \emph{Nature} \textbf{1998}, 395, 783-785.

\bibitem{TORRES}
Torres, J.A. and S\'{a}enze, J.J. \emph{Phys. Rev. Lett.}
\textbf{1996}, 77, 2245-2248.

\bibitem{OLESEN}
Olesen, L.; Brandbyge, M.; Sorensen, M.R.; Jacobsen, K.W.;
Laegsgaard, E.; Stensgaard, I.; Besenbacher, F. \emph{Phys. Rev.
Lett.} \textbf{1996}, 76, 1485-1488.

\bibitem{UNTIEDT}
Untiedt, C.; Caturla, M.J.; Calvo, M.R.; Palacios, J.J.; Segers, R.
C.; van Ruitenbeek, J.M. \emph{Phys. Rev. Lett.} \textbf{2007}, 98,
206801.

\bibitem{TAO2}
Huang, Z.; Chen, F.; Benett, P.A.; Tao, N.J. \emph{J. Am. Chem.
Soc.} \textbf{2007}, 129, 13225-13231.

\bibitem{TAO5}
Xu, B.; Xiao, X.; Tao, N.J. \emph{J. Am. Chem. Soc.} \textbf{2003},
125, 16164-16165.

\bibitem{RUBIO}
Rubio-Bollinger, G.; Bahn, S.R.; Agra\"it, N.; Jacobsen, K.W.;
Vieira, S. \emph{Phys. Rev. Lett.} \textbf{2001}, 87, 026101.

\bibitem{TAO6}
Xia, J.L.; Diez-Perez, I.; Tao, N.J., \emph{Nano Lett.}
\textbf{2008}, on-line, DOI: 10.1021/nl080857a.

\bibitem{GABOR}
M\'esz\'aros, G.; Li, C.; Pobelov, I.; Wandlowski, T.
\emph{Nanotechnology} \textbf{2007}, 18, 424004.

\bibitem{HAISS2}
Haiss, W.; van Zalinge, H.; Bethell, D.; Ulstrup, J.; Schriffin,
D.J.; Nichols, R.J. \emph{Faraday Discuss.} \textbf{2006}, 131,
253-264.

\bibitem{COSTA}
Costa-Kr\"amer, J.L. \emph{Phys. Rev. B.} \textbf{1997}, 55,
R4875-R4878.

\end{thebibliography}
\end{document}